%% file: arxiv.tex
\documentclass[11pt]{article}
\usepackage[a4paper,margin=2.5cm]{geometry}
\usepackage{amsmath,amssymb,amsfonts}
\usepackage{graphicx}
\usepackage{xcolor}
\usepackage[numbers,sort&compress]{natbib}
\usepackage[font=small,labelfont=bf]{caption}
\usepackage[colorlinks,allcolors=blue!50!black]{hyperref}

\newcommand{\SMref}[1]{Appendix~\ref{#1}}
\newcommand{\Letter}{the main text}
\newcommand{\secref}[1]{Appendix~\ref{#1}}
\newcommand{\Secref}[1]{Appendix~\ref{#1}}
\newcommand{\Lref}[1]{\ref{#1}}
\newcommand{\Leqref}[1]{\eqref{#1}}
\newcommand{\introheading}{\section{Introduction}}
\newcommand{\head}[1]{\section{#1}}
\newenvironment{acknowledgments}{\section*{Acknowledgments}}{}

\title{Microswimmers lose by swimming selfishly}
\author{Albert Alonso\\[2pt]
  \small Department of Computer Science, Aalborg University, Copenhagen, Denmark\\
  \small\texttt{aadlf@cs.aau.dk}}
\date{}

\begin{document}
\maketitle

\begin{abstract}
\input{content/abstract}
\end{abstract}

\input{content/body}

\clearpage
\appendix
% Appendix headings read "Appendix A. Title".
\makeatletter\renewcommand{\@seccntformat}[1]{Appendix~\csname the#1\endcsname.\quad}\makeatother
\counterwithin*{equation}{section}
\renewcommand{\theequation}{\thesection\arabic{equation}}
\input{content/supplement}

\bibliographystyle{unsrtnat}
\bibliography{refs}

\end{document}

%% file: content/abstract.tex
At low Reynolds number, microswimmers are coupled by the flows they create, even when they are apart.
The best way for each swimmer to push itself forward then depends on its neighbors, turning swimming together into a game.
To study this, we optimize the shape changes of soft swimmers, alone and in pairs, with a differentiable Stokes solver.
From random initial membrane motions, a lone swimmer converges to a traveling membrane wave.
In a pair, the stroke of a swimmer affects its neighbor far more than the neighbor affects its own best stroke.
For swimmers in line, this influence decays as $1/\ell$ with their separation $\ell$, whereas the incentive to respond decays as $1/\ell^2$.
Selfish swimmers, which maximize their own payoff, respond only to the weak incentive.
The cooperative gain they miss decays only as $1/\ell^2$, and their selfish adaptations leave the pair worse off than if both had kept their solitary strokes.
Related swimmers, which also value the payoff of their neighbor, recover this gain, whereas competitors pay to hinder each other, both following a law that we derive analytically.
Hydrodynamics alone thus makes swimming in company a social dilemma, whose outcome depends on how much the swimmers value each other.

%% file: content/body.tex
% ---------------------------------------------------------------------------
% Intro
% ---------------------------------------------------------------------------
\introheading
Microorganisms rarely swim alone.
Bacteria, amoeboid cells, sperm and algae move in groups and colonies~\cite{ben2000cooperative,rappel1999self,elgeti2015physics,goldstein2015green}.
In such groups, cells interact through the fluid they swim in, and at the scale of a cell, viscosity dominates over inertia (low Reynolds number)~\cite{purcell1977life,lauga2009hydrodynamics}, meaning that the flow each stroke sets up appears instantly, decays only slowly with distance, and is very much felt by neighboring cells.
Through these flows, swimmers with fixed strokes attract, repel and lock their phases~\cite{taylor1951analysis,ishikawa2006hydrodynamic,elfring2009hydrodynamic,uchida2011generic}, and the coupling can have strong consequences, such as making colonies feed worse than single cells~\cite{kirkegaard2016filter}.

Even at larger scales, where inertia dominates, fish can learn to exploit the wake of a leader~\cite{verma2018efficient}.
Whether the coupling also changes the stroke of self-promoting cells in the more strongly coupled regime of low Reynolds number is still unknown.
Nevertheless, the strength of these interactions suggests that it may, and if the best stroke for one cell depends on the strokes of the others, swimming in a group becomes a game.
According to evolutionary game theory~\cite{maynardsmith1973logic}, natural selection acting on each cell then favors a Nash equilibrium~\cite{nash1950equilibrium}, a set of strokes in which no cell can improve its own payoff by changing only its own stroke.
At such an equilibrium every cell does the best it can for itself, yet the group may swim worse than if the cells had chosen their strokes together.
Whether this matters or not depends on how much worse the equilibrium is than the cooperative optimum, a gap measured by the price of anarchy (PoA)~\cite{koutsoupias1999worst}.

For a single swimmer, the question of how best to swim is well understood.
Amoeboid cells such as \emph{Dictyostelium} and neutrophils can swim by changing their shape alone~\cite{barry2010dictyostelium,farutin2013amoeboid}, and for small deformations the optimal stroke of such a body is known to be a traveling wave~\cite{shapere1989efficiencies}.

Here, we ask what happens to this optimum when the swimmer has company.
To avoid building the answer into the model, we assume no particular form for the stroke and instead optimize a general Fourier representation of the membrane shape over time, starting from random strokes that we improve by gradient-based optimization~\cite{byrd1995limited}.
By making the Stokes solver differentiable~\cite{alonso2024mind}, we not only optimize the shape but also obtain the gradient of the payoff of each swimmer with respect to the strokes of its neighbors, which measures how strongly the swimmers influence each other.

We find that the flows between swimmers create a social dilemma.
The effect of a swimmer on its neighbor is much larger than the reason the neighbor gives it to modify its own stroke, which creates an interesting problem for the cells.
Even when they selfishly optimize their own movement, the effect of their changes on each other leaves both in a worse position than if they had not adapted at all.
It is therefore only when the cells help each other, as natural selection favors among related cells, that swimmers collect most of the benefit of swimming together.
Notably, the same strong effect means that a swimmer can disrupt the movement of its neighbor, but only by hindering its own performance.

\head{Model}Let each swimmer be a closed membrane in the plane, given by its radius in the body frame,
\begin{equation}
    r(s,t) = R(t)\,e^{\hat r(s,t)},
    \label{eq:radius}
\end{equation}
where $s$ is the polar angle measured from the body axis, the exponential keeps the radius positive for any deformation, and the scale $R(t)$ keeps the enclosed area constant, as for an incompressible cell.
We expand the deformation in a Fourier series in angle and time,
\begin{equation}
    \hat r(s,t) = \sum_{n,m} c_{nm} \cos(ns)\, e^{im\omega_0 t},
    \label{eq:fourier}
\end{equation}
with fundamental frequency $\omega_0 = 2\pi/T$ over the period $T$, $n \in \{0, 2, 3, \dots, N\}$, $|m| \le M$, and $c_{n,-m} = c^\ast_{nm}$ for a real radius.
Keeping only the terms in $\cos(ns)$ makes the shape mirror-symmetric about the body axis, which prevents the cell from turning.
While this enforced symmetry may seem restrictive, optimal strokes over one period can otherwise turn the cell, an artifact of scoring the displacement over a single period (see~\SMref{sec:single}).
Note that we leave out the mode $n=1$ to avoid moving the body off the polar origin.
We optimize the stroke of each cell, given here by its set of coefficients $\theta_i = \{c_{nm}\}$ for swimmer $i$.

The surrounding fluid, of viscosity $\mu$, obeys the two-dimensional Stokes equations, and a point $\mathbf x$ on swimmer $i$ moves with the membrane velocity $\mathbf u_i = \mathbf U_i + \Omega_i\,\hat{\mathbf z}\times(\mathbf x - \mathbf X_i) + \mathbf v_i$, where $\mathbf v_i$ is the deformation velocity set by the stroke, and $\mathbf U_i$ and $\Omega_i$ are the unknown translation and rotation about the centroid $\mathbf X_i$.
The membrane exerts a force density $\mathbf f_i$ on the fluid, and the no-slip condition requires the flow created by all swimmers to match the membrane velocity,
\begin{equation}
    \mathbf u_i(\mathbf x) = \sum_j \oint_j \mathbf G(\mathbf x - \mathbf y)\,\mathbf f_j(\mathbf y)\,ds_y ,
    \label{eq:bie}
\end{equation}
where $\mathbf G$ is the two-dimensional Stokeslet.
Each swimmer is force- and torque-free, $\oint_i \mathbf f_i\,ds = 0$ and $\oint_i (\mathbf x - \mathbf X_i)\times\mathbf f_i\,ds = 0$, which determines $\mathbf U_i$ and $\Omega_i$ at every instant.
Being force-free also avoids the Stokes paradox~\cite{happel1983low}, as the flow of a force-free swimmer decays at least as fast as $1/r$, even though the Stokeslet itself grows logarithmically with distance (see~\SMref{sec:multipole}).

We solve Eq.~\eqref{eq:bie} with the method of regularized Stokeslets~\cite{cortez2001method}, which places point forces on the membrane and spreads each of them over a small region of width $\sigma$ with the kernel
\begin{equation}
    \psi_\sigma(r) = \frac{3\sigma^3}{2\pi\,(r^2+\sigma^2)^{5/2}}.
    \label{eq:blob}
\end{equation}
We place the points at equal distances $\sigma$ along the membrane, where they move with the material as the cell deforms.

The energy a swimmer spends over a period is the work it does on the fluid,
\begin{equation}
    E_i = \int_0^T\!\!\oint_i \mathbf f_i\cdot\mathbf u_i\,ds\,dt
        = \int_0^T\!\!\oint_i \mathbf f_i\cdot\mathbf v_i\,ds\,dt.
    \label{eq:energy}
\end{equation}
With no inertia at low Reynolds number, this work is dissipated by viscosity instantly, and the dissipation in the fluid equals the work of all swimmers together.

Each swimmer aims to travel as far as possible for as little energy as possible, and we write its payoff over a period as the Lagrangian
\begin{equation}
    w_i = d_i - \lambda E_i ,
    \label{eq:payoff}
\end{equation}
where $d_i$ is the net displacement, $E_i$ the energy the swimmer spends, and $\lambda$ the Lagrange multiplier.

Cell membranes bend easily but hardly stretch~\cite{rawicz2000effect}, and as in models of amoeboid swimming~\cite{farutin2013amoeboid,ranganathan2018effect} we treat the membrane as inextensible and free to bend, with a perimeter $L_0$ a fraction $\epsilon$ longer than that of a circle of the same area.
A circle cannot change shape without stretching, and it is this excess membrane, which real cells keep in folds and other reservoirs~\cite{raucher1999characteristics}, that allows the cell to swim at all.
We use $\epsilon = 0.1$, a reduced area $4\pi A/L_0^2 \approx 0.83$.

Because Stokes flow has no intrinsic time scale, a stroke performed faster covers the same distance per beat but costs more energy. Thus, a larger $\lambda$ only makes the optimal stroke slower, which amounts to choosing the unit of time.
We use L-BFGS~\cite{byrd1995limited}, with gradients obtained by automatic differentiation through the full simulation~\cite{alonso2024mind}, to find the optimal stroke such that
\begin{equation}
    \theta^\ast = \arg\max_\theta \Bigl[\, w - \kappa \int_0^T \bigl(L(t) - L_0\bigr)^2 dt \,\Bigr],
    \label{eq:optimum}
\end{equation}
where the stiff quadratic penalty, of strength $\kappa$, holds the perimeter at $L_0$.

\begin{figure}[tb]
    \centering
    \includegraphics{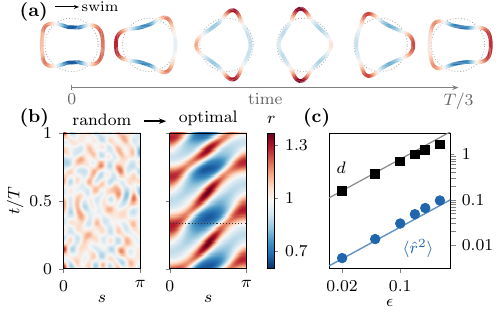}
    \caption{The optimal stroke of a lone swimmer.
    (a)~The membrane over a third of the period $T$, colored by its radius relative to the dotted circle.
    (b)~Membrane radius $r(s,t)$ from the front ($s=0$) to the back ($s=\pi$) over one period, before and after optimization. The dotted line marks the end of the window in (a).
    (c)~Mean-square deformation $\langle\hat r^2\rangle$ (circles) and displacement per period $d$ (squares) against the excess membrane $\epsilon$. The blue line is the prediction $\langle\hat r^2\rangle = 2\epsilon/\langle n^2-1\rangle$ (see~\SMref{sec:single}), with $\langle n^2-1\rangle \approx 8.0$ averaged over the scan, and the gray line has slope one.}
    \label{fig:single}
\end{figure}

% ---------------------------------------------------------------------------
% Lone swimmer
% ---------------------------------------------------------------------------
\head{A lone swimmer}We first optimize a single swimmer, starting from random coefficients $c_{nm}$.
In the optimal stroke [Fig.~\ref{fig:single}(a)], a wave of deformation travels from the front of the cell to the back on both sides at once and pushes the cell forward by nearly three quarters of its radius every period.
Notably, this wave emerges from a random initial motion [Fig.~\ref{fig:single}(b)], with different random starts also converging to the same stroke (see~\SMref{sec:single}).

The excess membrane sets the amplitude of this wave.
Displacement, dissipation, and with them the payoff all grow with the square of the deformation~\cite{shapere1989efficiencies,farutin2013amoeboid}, and the optimal stroke uses all the membrane it has, with a mean-square deformation that follows from the excess alone [Fig.~\ref{fig:single}(c) and~\SMref{sec:single}].

This stroke agrees with previous results, and at small amplitude it is the traveling wave found by Shapere and Wilczek~\cite{shapere1989efficiencies}, which is also the optimal stroke for slender and ciliated swimmers~\cite{alouges2019energy,michelin2010efficiency,guo2021optimal}.
Its chevron pattern resembles the protrusions observed on swimming \emph{Dictyostelium} cells~\cite{barry2010dictyostelium}, and the way its displacement grows with the excess membrane [Fig.~\ref{fig:single}(c)] agrees with models of amoeboid swimmers~\cite{farutin2013amoeboid}.
Here, we use $\theta^\ast$ to denote the optimal stroke of a single swimmer without neighbors.

% ---------------------------------------------------------------------------
% Pair: influence vs incentive
% ---------------------------------------------------------------------------
\head{Influence and incentive}We now place two swimmers in-line or lateral at a distance $\ell$, in units of the cell radius, with $5 \le \ell \le 20$ for the cells to stay at least two radii apart, and with both using their optimal stroke $\theta^\ast$.
We hold this arrangement fixed while computing the payoffs, as the separation drifts only over thousands of periods (see~\SMref{sec:numerics}).
To first order, the coupling between the two swimmers is described by two gradients, both taken with the two swimmers at $\theta^\ast$.
The first is how much swimmer $i$ could gain for itself by adapting its stroke to the neighbor,
\begin{equation}
    \mathbf h_i = \nabla_{\theta_i}\bigl[\,w_i(\theta_i,\theta_j) - w(\theta_i)\,\bigr],
    \label{eq:incentive}
\end{equation}
which we call the incentive.
Note that it vanishes for an isolated swimmer, because $\theta^\ast$ is the solitary optimum.
The second is how much the stroke of swimmer $i$ changes the payoff of its neighbor,
\begin{equation}
    \mathbf g_i = \nabla_{\theta_i} w_j(\theta_i,\theta_j),
    \label{eq:influence}
\end{equation}
which we call the influence.
As we show below, a selfish swimmer can exploit only its incentive, whereas cooperating swimmers also exploit their influence.
Differentiating the payoff of one swimmer through the solver gives both its incentive and the influence of its neighbor at once.

How these gradients decay with distance follows from the flow of a force-free swimmer (see~\SMref{sec:multipole}).
Far from the swimmer, its time-averaged flow is that of a force dipole of strength $S_0$ along the swimming axis $\hat{\mathbf e}$,
\begin{equation}
    \mathbf u(\mathbf x) = \frac{S_0\cos 2\alpha}{4\pi\mu\,|\mathbf x|}\,\hat{\mathbf x},
    \label{eq:dipole-main}
\end{equation}
where $\alpha$ is the angle between $\mathbf x$ and $\hat{\mathbf e}$.
An in-line neighbor ($\alpha = 0$, $|\mathbf x| = \ell$) is carried forward at speed $S_0/(4\pi\mu\ell)$, and a swimmer controls this push through $S_0$, which gives $|\mathbf g| \propto 1/\ell$.
The best stroke of a swimmer, in contrast, ignores a flow that is uniform across its body, which simply carries it along, and responds only to how the flow varies across the body, given by the derivative of Eq.~\eqref{eq:dipole-main}, and to its own flow reflected by the neighbor, which together give $|\mathbf h| \propto 1/\ell^2$.
For a lateral neighbor ($\alpha = \pi/2$), the flow of Eq.~\eqref{eq:dipole-main} points across the swimming direction, and the influence also falls to order $1/\ell^2$.

\begin{figure}[tb]
    \centering
    \includegraphics{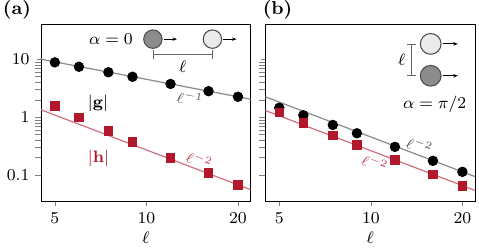}
    \caption{Influence and incentive at the solitary stroke. Influence $|\mathbf g|$ (circles) and incentive $|\mathbf h|$ (squares) of the dark cell against the separation $\ell$, for (a)~in-line and (b)~lateral pairs. Lines show the predicted scalings.}
    \label{fig:coupling}
\end{figure}

The simulations confirm these scalings [Fig.~\ref{fig:coupling}].
For in-line pairs, the influence exceeds the incentive at every separation, by a factor that grows from $6$ at $\ell = 5$ to $33$ at $\ell = 20$, and more than $98\%$ of the influence comes from the neighbor being carried along by the flow, rather than from changes in the viscous dissipation the two cells share.
For lateral pairs, both gradients decay as $\ell^{-2}$ and remain comparable at all separations, and selfish and cooperative swimmers should then change their strokes alike.
In-line swimmers that respond only to the incentive, in contrast, should leave most of the possible gain unused.

% ---------------------------------------------------------------------------
% Payoffs: price of anarchy
% ---------------------------------------------------------------------------
\head{Price of anarchy}We now let the two swimmers adapt their strokes, keeping their arrangement fixed.
A selfish swimmer maximizes its own payoff $w_i$ given the stroke of its neighbor, and we find the resulting Nash equilibrium, where neither cell can improve by a small change of its stroke, by letting the two swimmers take turns to play their best stroke, starting from $\theta^\ast$.
Cooperating swimmers instead maximize their total payoff $w_1 + w_2$, which we optimize jointly.
The price of anarchy (PoA), $\bar w_{\mathrm{coop}}/\bar w_{\mathrm{self}}$, where $\bar w$ is the payoff averaged over the two cells, is the factor by which a cooperating pair outperforms a selfish one, and $\mathrm{PoA}-1$ is the relative gain from cooperating.

\begin{figure}[tb]
    \centering
    \includegraphics{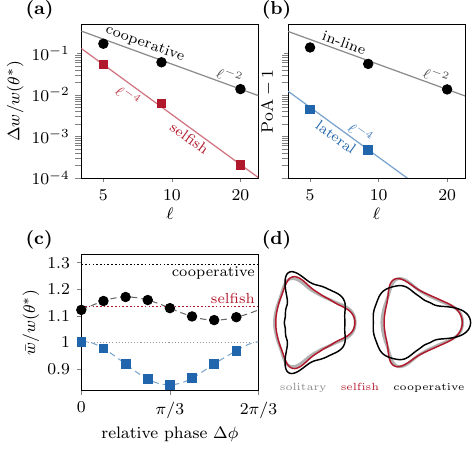}
    \caption{Selfish and cooperative strokes.
    (a)~Gain per cell from adapting the solitary stroke in in-line pairs, for cooperative cells (circles) and for a selfish cell whose neighbor keeps $\theta^\ast$ (squares).
    (b)~$\mathrm{PoA}-1$ for in-line (circles) and lateral (squares) pairs.
    (c)~Payoff per cell of a pair with solitary strokes against the relative phase at $\ell = 5$, for in-line (circles) and lateral (squares) pairs, with fits (dashed). Dotted lines mark lone cells and the selfish and cooperative in-line pairs.
    (d)~Shapes of the back and front cell of an in-line pair at $\ell = 5$, swimming to the right, drawn closer than their separation and with deviations from the solitary shape magnified three times.
    In (a) and (b), lines show the predicted scalings.}
    \label{fig:game}
\end{figure}

To see how these quantities depend on distance, we expand the payoffs around $\theta^\ast$ (see~\SMref{sec:game}).
Because $\theta^\ast$ maximizes the solitary payoff, a change of its own stroke alters the payoff of a swimmer only at second order, whereas the coupling to the neighbor contributes at first order through the two gradients.
If the two swimmers change their strokes by small amounts $\delta_i$ and $\delta_j$, the payoff of swimmer $i$ becomes
\begin{equation}
    w_i \approx w_i(\theta^\ast,\theta^\ast) - \tfrac12\,\delta_i^\top Q\,\delta_i + \mathbf h_i^\top\delta_i + \mathbf g_j^\top\delta_j ,
    \label{eq:expansion-main}
\end{equation}
where $w_i(\theta^\ast,\theta^\ast)$ is its payoff when both use $\theta^\ast$, and $Q = -\nabla^2 w(\theta^\ast)$ is the curvature of the solitary payoff at its optimum.
The three corrections are the cost of leaving the solitary optimum, the gain from changing its own stroke, and the gain from the change of its neighbor.
A selfish swimmer maximizes $w_i$ alone, whereas cooperating swimmers maximize $w_i + w_j$, which also contains the gain $\mathbf g_i^\top\delta_i$ that swimmer $i$ gives its neighbor, and the two change their strokes by
\begin{equation}
    \begin{aligned}
    \delta_i &= Q^{-1}\mathbf h_i && \text{(selfish)},\\
    \delta_i &= Q^{-1}\bigl(\mathbf h_i + \mathbf g_i\bigr) && \text{(cooperative)}.
    \end{aligned}
    \label{eq:shift-main}
\end{equation}
Inserted into Eq.~\eqref{eq:expansion-main}, these strokes show that a selfish swimmer whose neighbor keeps $\theta^\ast$ gains $\tfrac12\mathbf h_i^\top Q^{-1}\mathbf h_i \sim |\mathbf h|^2$, whereas each cooperating swimmer gains about $\tfrac12\mathbf g_i^\top Q^{-1}\mathbf g_i \sim |\mathbf g|^2$, which sets $\mathrm{PoA}-1$.
When both swimmers adapt selfishly, each also feels the change of its neighbor through the term $\mathbf g_j^\top Q^{-1}\mathbf h_j \sim |\mathbf g||\mathbf h|$.
Because the influence exceeds the incentive, this side effect outweighs what a swimmer gains from its own change, and its sign decides whether selfish adaptation helps the pair at all.
In line, the influence falls off as $1/\ell$ and the incentive as $1/\ell^2$, and the gain from cooperating then decays as $\ell^{-2}$, while that of a selfish swimmer decays as $\ell^{-4}$ and is already $65$ times smaller at $\ell = 20$.
For lateral pairs, both fall off as $1/\ell^2$, and the gain from cooperating decays as $\ell^{-4}$.
The simulations follow these scalings [Figs.~\ref{fig:game}(a) and~\ref{fig:game}(b)], with cooperating pairs doing $14\%$ better than selfish ones at $\ell = 5$ and $1.4\%$ better at $\ell = 20$ in line, and only $0.4\%$ better at $\ell = 5$ for lateral pairs.
These values hold for mirror-symmetric strokes, which keep the cells swimming straight (see~\SMref{sec:single}).

\head{A social dilemma}Part of the benefit of swimming together comes from timing alone.
Even with their solitary strokes, two cells gain or lose depending on their relative phase $\Delta\phi$, as found for waving sheets~\cite{taylor1951analysis,elfring2009hydrodynamic}.
Note that since the optimal stroke beats three times per period (see~\SMref{sec:numerics}), this payoff repeats every $2\pi/3$ in $\Delta\phi$ [Fig.~\ref{fig:game}(c)].
Both cells are best off at the same relative phase, a delay of $\Delta\phi \approx 0.5$ rad for in-line pairs and none for lateral ones, which leaves them to disagree only on the shape of the stroke.

Adapting the shape selfishly does not even beat this simple strategy.
At $\ell = 5$, two in-line cells that keep their solitary strokes and only adjust their relative phase gain $17\%$ over two lone cells, whereas the selfish pair gains only $14\%$ and the cooperative pair $29\%$.
Each selfish cell improves its own payoff given the stroke of the other, yet together they end up worse off than if neither had changed its shape, much as in a prisoner's dilemma~\cite{axelrod1981evolution}.

To see which cell adapts and how, we compare the strokes of the in-line pair at $\ell = 5$ [Fig.~\ref{fig:game}(d)], measuring the change of a stroke by the relative change in its Fourier coefficients after removing any shift in its timing.
The selfish front cell changes its stroke by only $5\%$, whereas the back cell, which swims in the wake of the front one, adapts by $12\%$.
Cooperating cells both change their strokes by about $27\%$, and the front cell now adapts not for its own benefit but because its stroke shapes the flow in which its follower swims.
Each selfish change does gain [Fig.~\ref{fig:game}(a)], but it also alters the payoff of the other cell, and here this side effect harms both.

\head{Relatives and competitors}Cells in a clone or a colony share their genes, and as natural selection carries these genes on, it also favors the payoff of related cells, through the weighted sum $w_i + \rho\, w_j$, where the weight $\rho$ that a cell gives to its neighbor is the relatedness of the two cells~\cite{hamilton1964genetical}.
It runs from $-1$ for competitors, which gain when their neighbor loses, through zero for unrelated cells to one for clones~\cite{hamilton1970selfish}.
Thus, we can express Eq.~\eqref{eq:shift-main} as $\delta_i = Q^{-1}(\mathbf h_i + \rho\,\mathbf g_i)$, which interpolates between the selfish and cooperative strokes.
Inserting this stroke into Eq.~\eqref{eq:expansion-main}, we find that the payoff $\bar w_\rho$ of the pair, averaged over its two cells, differs from that of a selfish pair by
\begin{equation}
    \bar w_\rho - \bar w_0 = \bigl(2\rho - \rho^2\bigr)\,(\bar w_1 - \bar w_0) ,
    \label{eq:kin}
\end{equation}
where the cooperative gain $\bar w_1 - \bar w_0$, between cooperative and selfish pairs, is set by the influence alone (see~\SMref{sec:game}).
Here, the first term sets the help a swimmer receives from its neighbor, while the second sets what it gives up to help in return.
Since the incentive does not depend on $\rho$, it cancels in this difference.
From Eq.~\eqref{eq:kin} we see that clones ($\rho = 1$) recover the gain that selfish cells forgo, and that competitors ($\rho < 0$) lose more the more they focus on reducing the payoff of the other.

\begin{figure}[tb]
    \centering
    \includegraphics{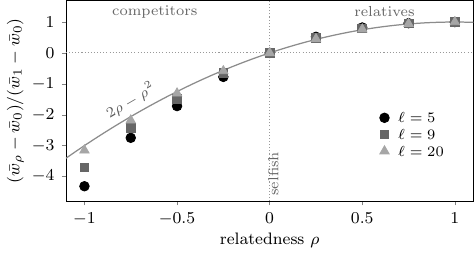}
    \caption{Relatives and competitors. Payoff per cell of an in-line pair against the relatedness $\rho$ of its cells, measured from that of a selfish pair in units of the cooperative gain $\bar w_1 - \bar w_0$, at three separations. The line is the prediction of Eq.~\eqref{eq:kin}.}
    \label{fig:kin}
\end{figure}

The simulations agree closely with the theoretical curve [Fig.~\ref{fig:kin}], with $\rho = 0$ and $\rho = 1$ on the curve by construction.
While the agreement is strong for relatives ($0 < \rho < 1$), the prediction does worse for competing cells at close range ($\ell = 5$), as these cells change their strokes so much that the expansion no longer holds.
We observe that the two cells of a pair gain almost exactly the same, within $2\%$, and the front--back asymmetry of the in-line pair plays little role here.

% ---------------------------------------------------------------------------
% Discussion
% ---------------------------------------------------------------------------
\head{Discussion}Hydrodynamic coupling turns swimming in company into a game, but a lopsided one.
The stroke of a swimmer strongly affects its neighbor, while the neighbor gives it little reason to change, and selfish swimmers, which respond only to this weak incentive, miss a gain set by their influence and leave each other worse off than if they had kept their solitary strokes.
Because the flow of a force-free swimmer decays slowly, as $1/\ell$ in two dimensions, this missed gain is long-ranged, and $\mathrm{PoA}-1$ decays only as $\ell^{-2}$ for swimmers in line.

In three dimensions the dipole flow decays as $1/\ell^2$, and the same argument gives $\mathrm{PoA}-1 \propto \ell^{-4}$ (see~\SMref{sec:multipole}), which confines the dilemma to close neighbors, where hydrodynamic interactions between real swimmers also stand out from noise~\cite{drescher2011fluid}.
In bacterial swarms and other crowded groups the cells are closer than our separations of $5$ to $20$ cell radii, and the near field will add to the effects described here.

The gain that selfish swimmers miss is available to cells that share their payoffs, as relatives do.
Related cells should then cooperate, whereas competing cells pay a cost to hinder each other, as in sperm competition~\cite{parker1970sperm}.
Natural examples are the clonal cells of a colony, or the sperm of deer mice, which swim faster in groups formed with sperm of the same male while competing with those of other males~\cite{fisher2010competition}.
Experimentally, we expect unrelated swimmers in line to adapt unequally, with the follower changing its stroke more than twice as much as the leader, whereas related swimmers should both change by similar amounts.

In larger groups the equilibria are harder to define, but as Stokes flow is linear, the influence of a cell on each of its neighbors adds up, and the gain that selfish cells leave unused can grow with the number of neighbors.
The same tension between the benefit of a cell and its effect on its neighbors has recently been proposed in chemical sensing, where a cell that absorbs a signal depletes it for the cells around it~\cite{kirkegaard2026observer}.
Microrobots trained by reinforcement learning on their own reward~\cite{colabrese2017flow,alonso2024learning} should likewise end in the selfish equilibrium unless they share it.
Even without any signal or contact, the fluid alone makes swimming a social act, and how a cell should swim depends not only on itself but on its company.

\begin{acknowledgments}
The author gratefully thanks Julius B. Kirkegaard for his conceptual input and many fruitful discussions, and Lukas H. Kristensen for his thorough feedback on this manuscript.
\end{acknowledgments}

%% file: content/supplement.tex
% ---------------------------------------------------------------------------
\section{The lone swimmer at small amplitude}
\label{sec:single}

\begin{figure}[h]
    \centering
    \includegraphics{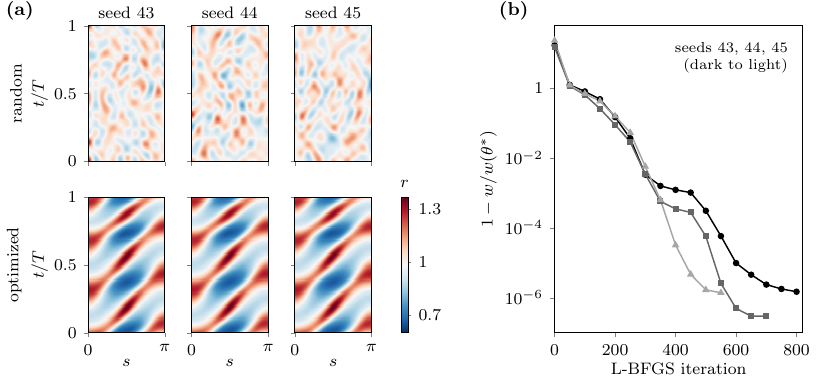}
    \caption{Random starts of the lone swimmer. (a)~Membrane radius $r(s,t)$, as in Fig.~\Lref{fig:single}(b) of \Letter{}, for three random initial strokes (top) and the strokes they converge to (bottom), shifted in time onto $\theta^\ast$. (b)~Distance of the payoff from that of $\theta^\ast$ during the optimization.}
    \label{fig:random}
\end{figure}

\emph{Random starts.}---Three optimizations from different random coefficients all converge to the optimal stroke $\theta^\ast$ [Fig.~\ref{fig:random}].
Within $800$ iterations their payoffs reach that of $\theta^\ast$ to within $2\times10^{-6}$, with two bulges ($n^\ast = 2$) and three beats per period ($m^\ast = 3$), and after removing an arbitrary shift in time their strokes correlate with $\theta^\ast$ above $0.999$.

\begin{figure}[h]
    \centering
    \includegraphics{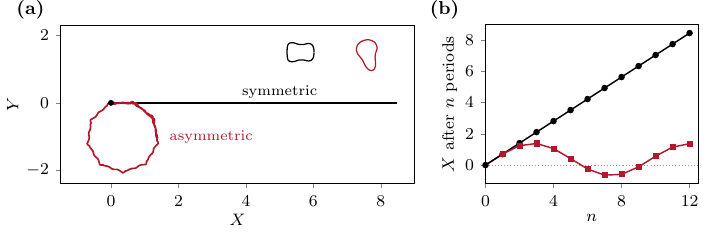}
    \caption{A lone swimmer without mirror symmetry. (a)~Path of the centroid over $12$ periods for the optimal stroke without the symmetry (red) and for the mirror-symmetric optimum $\theta^\ast$ (black), starting at the origin and heading along $x$, with lengths in cell radii. The outlines at the top show both shapes at $t = 0$. (b)~Displacement $X$ along the initial heading after $n$ periods.}
    \label{fig:asym}
\end{figure}

\emph{Without mirror symmetry.}---We repeated the lone optimization with all $\sin(ns)$ and $\cos(ns)$ terms, starting from a random stroke and letting the cell rotate.
The optimal stroke is still a wave with $n^\ast = 2$ that travels from front to back, but it becomes lopsided and turns the cell by about $0.7$ radian per period, and the cell swims in a circle with a radius of about one cell radius [Fig.~\ref{fig:asym}(a)].
Its payoff over one period is $29\%$ higher than that of the symmetric stroke, because a curved path still makes good progress along the initial heading within a single period [Fig.~\ref{fig:asym}(b)].
Over $12$ periods, however, it ends only $1.7$ radii from its start, against $8.4$ for the symmetric stroke.

Our objective, the displacement along the body axis over one period, is meaningful only for strokes that keep the heading fixed, which the mirror symmetry guarantees.
Finding whether an asymmetric stroke that swims straight can beat the symmetric one would require a longer horizon or a penalty on turning, which we do not pursue here.

\emph{Symmetry and lateral pairs.}---A lateral neighbor breaks the mirror symmetry of the surroundings of a swimmer, and the constraint could then bind for lateral pairs.
Without the symmetry and with the orientation of each cell held fixed, a cooperating lateral pair at $\ell = 5$ raises its payoff by $1.1\%$.
A lone cell under the same conditions, however, already gains about $1\%$ by breaking its symmetry, because a fixed orientation lets an asymmetric stroke score the forward speed of a cell that would in fact turn.
The gain of the pair thus comes mostly from this single-cell artifact, and a free cell would turn instead, as shown above.
The price of anarchy we report for lateral pairs applies to mirror-symmetric strokes, and the in-line results, where the neighbor does not break the symmetry, are unaffected.

\emph{Small-amplitude payoff.}---Write the coefficients of Eq.~\Leqref{eq:fourier} as $c_{nm} = a\,\hat c_{nm}$, where $\hat c$ sets the shape of the stroke and $a = \hat r_{\mathrm{rms}}$ is its amplitude, the root-mean-square value of $\hat r$ over the membrane and the stroke.
A circle that deforms slightly swims only at second order in $a$, because the first-order flow reverses when the stroke reverses.
The dissipation is also of second order.
Per unit time, the payoff is then
\begin{equation}
    \frac{w}{T} = a^2\Bigl[\omega\,\hat c^\dagger \mathsf A\,\hat c - \lambda\,\omega^2\,\hat c^\dagger \mathsf B\,\hat c\Bigr] + O(a^4),
    \label{eq:quadratic}
\end{equation}
with stroke frequency $\omega$ and two matrices that depend only on the circle.
The matrix $\mathsf A$ gives the swimming speed.
A circle translates only through products of neighboring modes, $n$ and $n\pm1$, so $\mathsf A$ couples only neighboring modes.
The matrix $\mathsf B$ gives the viscous dissipation.
The arguments below use only the fact that both are quadratic forms, and not their explicit entries, which can be computed along the lines of Ref.~\cite{shapere1989efficiencies}.

\emph{Unbounded amplitude.}---The bracket in Eq.~\eqref{eq:quadratic} is a downward parabola in $\omega$.
Its maximum is at $\omega^\ast = \hat c^\dagger\mathsf A\hat c/(2\lambda\,\hat c^\dagger\mathsf B\hat c)$, which leaves
\begin{equation}
    \frac{w}{T} = a^2\,\Lambda(\hat c),
    \qquad
    \Lambda(\hat c) = \frac{(\hat c^\dagger\mathsf A\,\hat c)^2}{4\lambda\,\hat c^\dagger\mathsf B\,\hat c} .
    \label{eq:lambda}
\end{equation}
Swimming and dissipation are both quadratic in the amplitude, which gives the beat frequency a well-defined optimum but leaves the amplitude free.
Since $\Lambda > 0$, the payoff $a^2\Lambda$ keeps growing with the amplitude, and only the membrane can fix it.
Optimal strokes in Stokes flow always need a limit on the deformation, such as a fixed area or a bounded amplitude~\cite{shapere1989efficiencies,avron2004optimal,michelin2010efficiency}, and without one our simulations indeed run away.

\emph{Amplitude from the excess membrane.}---A membrane that cannot stretch keeps its length $L_0 = 2\pi r_0(1+\epsilon)$, where $r_0$ is the radius of a circle with the same area as the cell.
For the shape of Eq.~\Leqref{eq:radius} at fixed area, expanding the perimeter to second order in $\hat r$ gives the excess length
\begin{equation}
    \epsilon = \frac{L}{2\pi r_0} - 1 \approx \frac{1}{4\pi}\oint \bigl(\hat r_s^2 - \hat r^2\bigr)\, ds ,
    \label{eq:excess}
\end{equation}
where $\hat r_s = \partial_s \hat r$.
A deformation of mode $n$ has $\hat r_s^2$ on average $n^2$ times larger than $\hat r^2$, so each mode lengthens the perimeter in proportion to $n^2-1$.
Averaging over the stroke, the constraint $L = L_0$ becomes
\begin{equation}
    a^2 = \frac{2\epsilon}{\langle n^2-1\rangle},
    \label{eq:amplitude}
\end{equation}
where $\langle n^2-1\rangle$ is the average of $n^2-1$ weighted by the power of the stroke in each mode, $\hat c^\dagger\mathsf L\,\hat c$ with $\mathsf L$ diagonal, so that the amplitude follows from the geometry of the membrane alone.
For a single traveling wave $A_n\cos(ns-\omega t)$, whose mean square is $A_n^2/2$, Eq.~\eqref{eq:amplitude} gives $A_n = 2\sqrt{\epsilon/(n^2-1)}$.
Because the swimmer would deform further if it could, the constraint pushes back, and the membrane tension is positive.
For the optimal stroke at $\epsilon = 0.1$, the measured $\langle n^2-1\rangle = 7.7$ gives $a = 0.161$, against the measured $0.173$.
The difference comes from the terms beyond second order in the expansion of the perimeter, which grow with the amplitude, as Fig.~\Lref{fig:single}(c) of \Letter{} shows.

\emph{The wavelength.}---Inserting Eq.~\eqref{eq:amplitude} into Eq.~\eqref{eq:lambda} gives the payoff of a stroke shape $\hat c$,
\begin{equation}
    \frac{w}{T} = 2\epsilon\,\frac{\Lambda(\hat c)}{\hat c^\dagger\mathsf L\,\hat c} .
    \label{eq:wavelength}
\end{equation}
The optimal shape maximizes this ratio, and its dominant mode $n^\ast$ sets the wavelength.
Short waves use more membrane, since $\hat c^\dagger\mathsf L\,\hat c$ grows as $n^2-1$, and at fixed excess their amplitude falls as $1/n$.
In the simulations the benefit $\Lambda$ does not grow fast enough with $n$ to compensate, and the optimum is the longest wave allowed, $n^\ast = 2$, at every excess membrane we tried.

% ---------------------------------------------------------------------------
\section{How the influence and the incentive decay with distance}
\label{sec:multipole}

\emph{The far field of a swimmer.}---Let $\mathbf f(\mathbf y)$ be the force per unit length that a swimmer centered at the origin exerts on the fluid.
In two dimensions, the flow it creates is
\begin{equation}
    u_i(\mathbf x) = \oint \mathcal G_{ij}(\mathbf x - \mathbf y)\, f_j(\mathbf y)\, ds_y,
    \qquad
    \mathcal G_{ij}(\mathbf x) = \frac{1}{4\pi\mu}\Bigl(-\delta_{ij}\ln|\mathbf x| + \frac{x_i x_j}{|\mathbf x|^2}\Bigr),
    \label{eq:stokeslet2d}
\end{equation}
where $\mathcal G$ is the two-dimensional Stokeslet.
Far from the swimmer, $|\mathbf y| \ll |\mathbf x|$, and we expand $\mathcal G_{ij}(\mathbf x - \mathbf y) \approx \mathcal G_{ij}(\mathbf x) - y_k\, \partial_k \mathcal G_{ij}(\mathbf x)$.
The first term multiplies the net force, which is zero, and the flow starts with the second term,
\begin{equation}
    u_i(\mathbf x) \approx -D_{jk}\, \partial_k \mathcal G_{ij}(\mathbf x),
    \qquad
    D_{jk} = \oint f_j\, y_k\, ds_y ,
    \label{eq:dipole-moment}
\end{equation}
where $D$ is the force dipole of the swimmer.
The antisymmetric part of $D$ is the torque, which is zero.
Its trace produces no flow, because the Stokeslet is divergence-free, $\partial_j \mathcal G_{ij} = 0$.
Only the symmetric, traceless part $\mathbf S$ remains.
The derivative of the Stokeslet is
\begin{equation}
    \partial_k \mathcal G_{ij} = \frac{1}{4\pi\mu}\Bigl(\frac{-\delta_{ij} x_k + \delta_{ik} x_j + \delta_{jk} x_i}{|\mathbf x|^2} - \frac{2\, x_i x_j x_k}{|\mathbf x|^4}\Bigr).
\end{equation}
Contracted with a symmetric, traceless $\mathbf S$, the first three terms cancel, and only the last one survives.
Averaged over a stroke period, the far field is
\begin{equation}
    \mathbf u(\mathbf x) = \frac{\hat{\mathbf x}\cdot\bar{\mathbf S}\cdot\hat{\mathbf x}}{2\pi\mu\,|\mathbf x|}\;\hat{\mathbf x},
    \qquad
    \hat{\mathbf x} = \frac{\mathbf x}{|\mathbf x|},
    \label{eq:dipole}
\end{equation}
where $\bar{\mathbf S}$ is the time-averaged dipole.
It decays as $1/|\mathbf x|$, more slowly than the $1/|\mathbf x|^2$ of three dimensions.

For a swimmer that is mirror-symmetric about its axis $\hat{\mathbf e}$, the dipole must have the same symmetry, which gives $\bar{\mathbf S} = S_0\,(\hat{\mathbf e}\hat{\mathbf e} - \tfrac12\mathbf I)$ with a single strength $S_0$.
Writing $\alpha$ for the angle between $\hat{\mathbf x}$ and $\hat{\mathbf e}$, Eq.~\eqref{eq:dipole} becomes
\begin{equation}
    \mathbf u = \frac{S_0 \cos 2\alpha}{4\pi\mu\,|\mathbf x|}\;\hat{\mathbf x}.
    \label{eq:radial}
\end{equation}
The flow points directly towards or away from the swimmer.

\emph{The influence.}---The neighbor sits at distance $\ell$ and angle $\alpha$.
Only the part of the flow along the swimming direction changes its payoff, and Eq.~\eqref{eq:radial} gives
\begin{equation}
    \mathbf u\cdot\hat{\mathbf e} = \frac{S_0 \cos 2\alpha \cos\alpha}{4\pi\mu\,\ell}.
    \label{eq:push}
\end{equation}
Over the horizon $T$, the neighbor is carried a distance $A = T\,\mathbf u\cdot\hat{\mathbf e}$.
The stroke of the swimmer enters only through its dipole strength $S_0$, which gives the influence $\mathbf g = \nabla_\theta A \propto \nabla_\theta S_0 / \ell$.

For in-line pairs ($\alpha = 0$), $\mathbf g$ is of order $1/\ell$.
For lateral pairs ($\alpha = \pi/2$), the flow at the neighbor is perpendicular to the swimming direction, and the $1/\ell$ term vanishes.
The influence then starts at the next order, $1/\ell^2$, as measured for lateral pairs.
The same argument in three dimensions, where the dipole flow decays as $1/\ell^2$, gives $\mathbf g \propto 1/\ell^2$ and $\mathrm{PoA} - 1 \propto 1/\ell^4$ for in-line pairs.

\emph{The incentive.}---The incentive $\mathbf h$ is the part of the interaction that depends on the stroke of the swimmer itself.
A uniform flow cannot contribute to it, because in Stokes flow a swimmer in a uniformly moving fluid behaves exactly as in a fluid at rest and is carried along with the same best stroke.
The best stroke can change only through the variation of the flow of the neighbor across the body of the swimmer, given by the gradient of Eq.~\eqref{eq:dipole}, and through the flow of the swimmer reflected by the body of the neighbor.
The variation is of order $1/\ell^2$ and acts on the shape changes of the swimmer, on its small oscillations in position, and on the shared dissipation.
The reflected flow reaches the neighbor with a gradient of order $1/\ell^2$, the neighbor responds with a dipole of that order, and its flow travels back with a further factor $1/\ell$, which makes the reflected flow of order $1/\ell^3$.
The incentive is thus of order $1/\ell^2$, a factor $1/\ell$ below the influence of in-line pairs and comparable to that of lateral pairs.
The simulations confirm that the incentive decays as $1/\ell^2$ for both in-line and lateral pairs [Fig.~\Lref{fig:coupling} of \Letter{}].

% ---------------------------------------------------------------------------
\section{Numerical parameters}
\label{sec:numerics}

Table~\ref{tab:numerics} lists the parameters of all simulations.
Energies are in units where $4\pi\mu = 1$, and lengths in units of the radius of a circle with the same area as the cell.

\begin{table}[h]
\centering
\begin{tabular}{lll}
\hline\hline
Quantity & Symbol & Value \\
\hline
Spatial modes & $N$ & 12 ($n = 0, 2, 3, \dots, 12$) \\
Temporal modes & $M$ & 8 \\
Basis frequency and period & $\omega_0$, $T$ & 1, $2\pi$ \\
Optimization horizon & & one period $T$ \\
Viscous weight & $\lambda$ & 0.0429 \\
Excess membrane & $\epsilon$ & 0.1 (varied in Fig.~\Lref{fig:single}(c) of \Letter) \\
Points on the membrane & & 192, equally spaced along the membrane \\
Kernel width & $\sigma$ & spacing between points \\
Time step & $\Delta t$ & $\approx 0.05$ \\
Perimeter penalty stiffness & $\kappa$ & 10 (perimeter within 0.2\% of $L_0$) \\
Optimizer & & L-BFGS, memory 100, line-search step $\le 0.1$ \\
\hline\hline
\end{tabular}
\caption{Parameters of the simulations.}
\label{tab:numerics}
\end{table}

\emph{The choice of $\lambda$.}---Because Stokes flow is rate independent, playing a stroke $k$ times faster multiplies its distance per period by $k$ and its dissipation by $k^2$.
The payoff of a given shape, $k\,d - \lambda k^2 E$, is largest at a beat frequency set by $\lambda$, so $\lambda$ only fixes the unit of time.
Because our strokes repeat after one period $T$ of the Fourier basis, the number of beats per period is a whole number.
We choose $\lambda = 0.0429$, for which the optimal beat frequency of the optimal shape falls at three beats per period.
The optimal stroke then has its dominant temporal harmonic at $m^\ast = 3$, and the highest harmonic $m = M$ carries only $0.3\%$ of its power.

\emph{Horizon.}---A lone swimmer repeats its stroke exactly, and optimizing over one period or three periods gives the same stroke, with payoffs per period that agree to $2\times10^{-5}$.
We thus optimize over one period.

\emph{Drift of the arrangement.}---In \Letter{} each pair is held at a fixed separation and angle.
Free swimmers would slowly move apart or together, because the two cells do not swim at exactly the same speed.
Table~\ref{tab:drift} gives the difference in distance per period $\Delta U$ between the back and front cell of in-line pairs, or between the two cells of lateral pairs, and the number of periods it would take to close the gap between them.
The fastest drift, about $1.5\times10^{4}$ periods, occurs for selfish in-line pairs at $\ell = 5$.
The arrangement is quasi-static on the time scale of a period.

\begin{table}[h]
\centering
\begin{tabular}{llll}
\hline\hline
Pair & solitary & selfish & cooperative \\
\hline
in-line, $\ell = 5$ & $9.2\times10^{-5}$ ($2.9\times10^{4}$) & $1.6\times10^{-4}$ ($1.5\times10^{4}$) & $8.9\times10^{-5}$ ($2.7\times10^{4}$) \\
in-line, $\ell = 9$ & $4.6\times10^{-5}$ ($1.5\times10^{5}$) & $3.6\times10^{-5}$ ($1.8\times10^{5}$) & $2.0\times10^{-5}$ ($3.2\times10^{5}$) \\
in-line, $\ell = 20$ & $2.0\times10^{-5}$ ($8.9\times10^{5}$) & $8.9\times10^{-6}$ ($2.0\times10^{6}$) & $4.2\times10^{-5}$ ($4.2\times10^{5}$) \\
lateral, $\ell = 5$ & 0 (symmetry) & $1.8\times10^{-6}$ ($1.3\times10^{6}$) & $2.8\times10^{-6}$ ($8.1\times10^{5}$) \\
lateral, $\ell = 9$ & 0 (symmetry) & $6.5\times10^{-6}$ ($9.6\times10^{5}$) & $5.8\times10^{-7}$ ($1.1\times10^{7}$) \\
\hline\hline
\end{tabular}
\caption{Difference in distance per period $|\Delta U|$ between the two cells of a pair, and in parentheses the number of periods needed to close the gap between them, for the solitary, selfish and cooperative strokes. The cells travel about $0.7$ radii per period.}
\label{tab:drift}
\end{table}

\emph{Resolution.}---The regularized Stokeslets converge to first order in the spacing between points.
Re-evaluating a lone optimum with $96$, $192$ and $384$ points, the differences in distance and payoff halve at each refinement.
For the reference stroke, doubling the points from $192$ to $384$ raises the payoff of a lone swimmer from $0.372$ to $0.412$, so the production values are about $10\%$ low.
Ratios are much less sensitive.
Re-evaluating the in-line strokes at $\ell = 5$ with $384$ points raises the total payoffs by $9$--$10\%$ but changes the price of anarchy by only $4\%$, from $\mathrm{PoA}-1 = 0.139$ to $0.133$, and the cooperative gain per cell by $7\%$.
The small selfish gain, a difference of close numbers, changes by $30\%$.
Solitary strokes at their best relative phase still beat the selfish equilibrium on the finer mesh.
Because these strokes were optimized at $192$ points, this tests the evaluation rather than equilibria re-optimized on the finer mesh.

% ---------------------------------------------------------------------------
\section{Selfish and cooperative strokes near the solitary optimum}
\label{sec:game}

We treat the neighbor as a small perturbation of the solitary problem, which gives the strokes of selfish, cooperative and related swimmers, and their payoffs, in closed form.

\emph{Setting.}---Two identical swimmers are at a distance $\ell$, and $\alpha$ is the angle between the swimming axis and the line joining them.
Swimmer $i$ uses the stroke $\theta_i$, which is its set of Fourier coefficients $c_{nm}$, and its payoff $w_i$ is Eq.~\Leqref{eq:payoff} of \Letter{}, evaluated over a whole number of periods.
We hold the arrangement fixed, with constant $\ell$ and $\alpha$ over the stroke, and evaluate the payoff of each swimmer at that arrangement.

\emph{Splitting the payoff.}---Because Stokes flow is linear, to leading order swimmer $i$ moves with its solitary velocity plus the velocity of the flow that swimmer $j$ creates at its position.
This splits the payoff of swimmer $i$ into three parts,
\begin{equation}
    w_i(\theta_i, \theta_j) = w(\theta_i) + A_{j}(\theta_j) + C_i(\theta_i, \theta_j).
    \label{eq:split}
\end{equation}
The first part, $w(\theta_i)$, is the payoff the swimmer would have alone.

The second part, $A_j$, is how far swimmer $i$ is carried by the time-averaged flow of swimmer $j$.
It depends only on the stroke of swimmer $j$, hence the label, because a flow that is uniform over the size of a swimmer carries it along, whatever stroke it uses.
\Secref{sec:multipole} shows that $A_j$ decays as $1/\ell$ in two dimensions.

The third part, $C_i$, contains everything that depends on both strokes at once, namely the viscous dissipation the two swimmers share, the coupling between the shape changes of swimmer $i$ and the gradients of the flow of swimmer $j$, and the flow of swimmer $i$ reflected back to it by the body of swimmer $j$.
All three decay at least as fast as $1/\ell^2$, as shown in \secref{sec:multipole}.
When the swimmers are not too close, $C_i$ is thus small compared with $A_j$.

\emph{Two gradients.}---Let $\theta^\ast$ be the solitary optimum, the stroke that maximizes $w$.
Near it, the solitary payoff is a downward parabola with curvature matrix $Q = -\nabla^2 w(\theta^\ast)$.
Because a shift of the stroke in time leaves the solitary payoff unchanged, $Q$ is zero along this direction, and $Q^{-1}$ denotes its pseudoinverse, taken on the other directions.
Two gradients, both taken with both swimmers at $\theta^\ast$, describe the interaction.
The \emph{influence}
\begin{equation}
    \mathbf g_i = \nabla_{\theta_i} A_i
    \label{eq:g}
\end{equation}
measures how much swimmer $i$ can change how far its neighbor is carried, by changing its own stroke.
It is the leading part of the gradient $\nabla_{\theta_i} w_j$ used in \Letter{}, which also contains the smaller term $\nabla_{\theta_i} C_j$.
The \emph{incentive}
\begin{equation}
    \mathbf h_i = \nabla_{\theta_i} C_i
    \label{eq:h}
\end{equation}
measures how much swimmer $i$ can gain for itself from the presence of the neighbor, by changing its own stroke.

\emph{The payoff near the solitary optimum.}---Let both swimmers change their strokes slightly, to $\theta_i = \theta^\ast + \delta_i$.
Expanding each part of Eq.~\eqref{eq:split} to the lowest order that matters gives
\begin{equation}
    w_i \approx w_i(\theta^\ast,\theta^\ast) - \tfrac12\, \delta_i^\top Q\, \delta_i + \mathbf h_i^\top \delta_i + \mathbf g_j^\top \delta_j ,
    \label{eq:expansion}
\end{equation}
where $w_i(\theta^\ast,\theta^\ast)$ is the payoff when both use $\theta^\ast$.
The first term has no linear part, because $\theta^\ast$ is the solitary optimum.
The quadratic terms of $A_j$ and $C_i$ are smaller than the ones kept and are dropped.
The remaining terms are the loss from leaving the solitary optimum, the gain of swimmer $i$ through the coupling from its own change, and the gain it receives from the change of its neighbor.

\emph{Selfish, cooperative and related swimmers.}---Following Hamilton~\cite{hamilton1964genetical}, swimmer $i$ maximizes
\begin{equation}
    \tilde w_i = w_i + \rho\, w_j ,
    \label{eq:hamilton}
\end{equation}
where the relatedness $\rho$ weighs the payoff of the neighbor.
Selfish swimmers have $\rho = 0$.
Cooperative swimmers, which maximize the total payoff, have $\rho = 1$.
Competing swimmers, which gain from the loss of the neighbor, have $\rho < 0$.
Swimmer $i$ controls only $\delta_i$.
By Eq.~\eqref{eq:expansion}, $\delta_i$ enters its own payoff through the first two terms and the payoff of the neighbor through $\mathbf g_i^\top \delta_i$.
The optimum satisfies
\begin{equation}
    \nabla_{\delta_i} \tilde w_i = -Q\,\delta_i + \mathbf h_i + \rho\, \mathbf g_i = 0 ,
\end{equation}
which gives the new stroke
\begin{equation}
    \delta_i = Q^{-1}\bigl(\mathbf h_i + \rho\, \mathbf g_i\bigr).
    \label{eq:shift}
\end{equation}
A selfish swimmer ($\rho = 0$) moves away from its solitary stroke by $Q^{-1} \mathbf h_i$, which is only as large as the small incentive.
A cooperative swimmer ($\rho = 1$) moves by $Q^{-1}(\mathbf h_i + \mathbf g_i)$, which is dominated by $Q^{-1} \mathbf g_i$ and is as large as the influence.
Because $|\mathbf h| < |\mathbf g|$, and increasingly so with distance for in-line pairs, the selfish stroke moves less than the cooperative one.

\emph{How much the payoff changes.}---Inserting Eq.~\eqref{eq:shift} into Eq.~\eqref{eq:expansion} and collecting powers of $\rho$, the terms linear in $\rho$ that involve $\mathbf h_i$ cancel, and the payoff changes by
\begin{equation}
    \Delta w_i(\rho) \approx \Delta w_i(0) + \rho\, \mathbf g_j^\top Q^{-1} \mathbf g_j - \tfrac12 \rho^2\, \mathbf g_i^\top Q^{-1} \mathbf g_i ,
    \qquad
    \Delta w_i(0) = \tfrac12\, \mathbf h_i^\top Q^{-1} \mathbf h_i + \mathbf g_j^\top Q^{-1} \mathbf h_j .
    \label{eq:parabola}
\end{equation}
The selfish change $\Delta w_i(0)$ involves the incentive and does not depend on $\rho$, and only shifts the parabola.
Since $\mathbf h$ is much smaller than $\mathbf g$, this offset is small compared with the terms in $\rho$.
Averaged over the two swimmers, the payoff $\bar w_\rho$ of the pair satisfies
\begin{equation}
    \bar w_\rho - \bar w_0 \approx \bigl(2\rho - \rho^2\bigr)\,(\bar w_1 - \bar w_0),
    \qquad
    \bar w_1 - \bar w_0 \approx \tfrac14\bigl(\mathbf g_1^\top Q^{-1} \mathbf g_1 + \mathbf g_2^\top Q^{-1} \mathbf g_2\bigr),
    \label{eq:parabola-sym}
\end{equation}
which holds even when the two swimmers are not in equivalent positions, as for in-line pairs, and for any relatedness $-1 \le \rho \le 1$.
Measured from the selfish pair, fully competing swimmers ($\rho = -1$) each lose three times the cooperative gain $\bar w_1 - \bar w_0$, because each one gives up payoff to harm the other.
The selfish gain itself is only of order $\mathbf g^\top Q^{-1} \mathbf h$.

\emph{Price of anarchy.}---The price of anarchy (PoA)~\cite{koutsoupias1999worst} is the total payoff of cooperative swimmers divided by that of selfish swimmers.
Each selfish swimmer earns $\bar w_0$ on average, and Eq.~\eqref{eq:parabola-sym} gives
\begin{equation}
    \mathrm{PoA} - 1 = \frac{\bar w_1 - \bar w_0}{\bar w_0} \approx \frac{\mathbf g_1^\top Q^{-1} \mathbf g_1 + \mathbf g_2^\top Q^{-1} \mathbf g_2}{4\,\bar w_0} .
    \label{eq:poa}
\end{equation}
The cooperative gain is quadratic in the influence.
Since the influence decays as $1/\ell$ for in-line pairs (\secref{sec:multipole}), $\mathrm{PoA} - 1$ decays as $1/\ell^2$.

\emph{Computing the strokes.}---The simulations do not use this expansion, which serves only to explain the results.
We find the selfish strokes by letting the two swimmers take turns to maximize their own payoff with L-BFGS, starting from $\theta^\ast$, and stop when neither can gain more than $10^{-4}\,w(\theta^\ast)$ by changing its stroke.
This converges within two or three rounds.
Because each best response is a local optimization, the result is a Nash equilibrium in the local sense that no swimmer can improve by a small change of its stroke.
We find the cooperative strokes by maximizing $w_1 + w_2$ over both strokes at once.
Restarting both optimizations from perturbed strokes gives the same total payoffs to within $10^{-3}$ of the payoff of a lone swimmer.

\emph{The relative phase.}---Shifting a stroke in time leaves the solitary payoff unchanged.
The phase of the stroke is thus a direction in which $w$ is flat, which the pseudoinverse in Eq.~\eqref{eq:shift} leaves out.
The time-averaged flow does not depend on phase either, so $A_j$ cannot set the phase.
Only the small cross term $C_i$ does, for selfish and cooperative swimmers alike.
Two swimmers can still lock their phases, but the phase they choose changes their payoffs only by an amount of order $C$.